\documentclass[10pt]{article} %
\usepackage[preprint]{tmlr}

\usepackage{amsmath,amsfonts,bm}

\def\eqref#1{equation~\ref{#1}}

\def\1{\bm{1}}

\DeclareMathAlphabet{\mathsfit}{\encodingdefault}{\sfdefault}{m}{sl}
\SetMathAlphabet{\mathsfit}{bold}{\encodingdefault}{\sfdefault}{bx}{n}

\usepackage{hyperref}
\usepackage{url}

\usepackage{booktabs}
\usepackage{graphicx}
\usepackage[ruled]{algorithm2e}
\usepackage{todonotes}
\title{Learning Best Response Policies in Dynamic Auctions via Deep Reinforcement Learning}
\author{\name Vinzenz Thoma\thanks{These authors contributed equally to this work.} \email vinzenz.thoma@ai.ethz.ch \\
\addr ETH Zurich \\ ETH AI Center \AND \name Michael Curry\footnotemark[1] \email curry@ifi.uzh.ch \\ \addr University of Zurich \\ Harvard University \\ ETH AI Center  \AND \name Niao He \email niao.he@inf.ethz.ch \\ \addr ETH Zurich  \AND \name  Sven Seuken \email seuken@ifi.uzh.ch \\ \addr University of Zurich \\ ETH AI Center}

\begin{document}

\maketitle

\begin{abstract}
    Many real-world auctions are dynamic processes, in which bidders interact and report information over multiple rounds with the auctioneer. The sequential decision making aspect paired with imperfect information renders analyzing the incentive properties of such auctions much more challenging than in the static case. It is clear that bidders often have incentives for manipulation, but the full scope of such strategies is not well-understood. We aim to develop a tool for better understanding the incentive properties in dynamic auctions by using reinforcement learning to learn the optimal strategic behavior for an auction participant. We frame the decision problem as a Markov Decision Process, show its relation to multi-task reinforcement learning and use a soft actor-critic algorithm with experience relabeling to best-respond against several known analytical equilibria as well as to find profitable deviations against exploitable bidder strategies. 
    \end{abstract}

\section{Introduction}
Auctions are widely used to buy and sell goods and services, from online advertisements to allocating resources such as spectrum rights or computing power \citep{Krishna2009Auction}. Ideally an auction mechanism is efficient and individually rational, i.e.~no bidder is worse off from participating. Moreover, it is highly desirable that the mechanism is incentive compatible, such that everyone is always best off by reporting truthfully. However, as it turns out, the only mechanism fulfilling the properties of efficiency, individual rationality, and incentive compatibility is the VCG mechanism, which is rarely used in practice because it is not robust to imperfect knowledge of bidder utilities, and also may generate low revenue, among other limitations. In contrast many mechanisms used in practice are not provably incentive compatible. At the same time, computing the equilibrium policies for the bidders might be computationally infeasible. Hence several works have instead proposed to study mechanisms at learning outcomes, i.e. study which properties mechanisms fulfil if bidders use different learning algorithms \citep{Daskalakis2022Learning,Weed2016Online,Guo2022NoRegret}. In this spirit, our work focuses on the problem of learning to bid optimally against a set of opponents. In particular, we would like to understand how great the incentives are in a particular auction for a bidder to deviate from desired behaviours such as bidding truthfully. For this problem, we propose to use reinforcement learning (RL) to learn optimal bidding strategies.

\section{Related Work}
The problem of learning in auctions has been studied widely. \cite{Daskalakis2022Learning,Guo2022NoRegret,Kolumbus2022Auctions} for example study auctions with no-regret learners. \cite{Nedelec2022Learning} present a larger survey on learning in repeated auctions. \cite{Jeunen2022Learning,Zhao2018Deep} specifically study applying RL to sponsored search auctions. \cite{Weed2016Online} take an online learning approach in repeated Vickrey auctions. \cite{Banchio2022Artificial} study how Q-learners can learn to collude in auctions. \cite{Bosshard2020Computing,Fichtl2022Computing} directly try to learn an equilibrium in one-round auctions. More generally \cite{Martin2022Finding,Chen2021Temporal} study the problem of learning equilibria in stochastic games, of which auctions are of course a prominent example.

While analytic solutions to dynamic auctions are spare, notable works include \cite{Krishna2009Auction,Kokott2019Beauty} who present equilibrium strategies for sequential sales and multi-round split-award auctions, which we will reproduce in this work.

A recent line of work considers iterative best-response algorithms for learning equilibria in combinatorial auctions~\citep{Bosshard2020Computing}. This was extended by \cite{Thoma2023Computing} to sequential auctions. Whiley they provide a verification for their computed equilibria, they have to make certain assumptions on the auction structure and the common knowledge of all bidders to perform game abstraction. In contrast, we do not make such assumptions. The abstraction is (if at all) done by the neural network. Similarly, \citet{Greenwald2012Approximating} try to learn best-responses in sequential auctions by framing the problem as an MDP. However, they first discretize the auctions and then solve the simplified game using linear programming, whereas we use deep RL directly on the full auction environment without abstraction.

\section{Problem formulation}
In this work, we study the problem of bidding in a dynamic auction, i.e. an auction format that spans across multiple consecutive rounds. Such formats have been previously studied in the context of spectrum or procurement auctions \citep{Shoham2008Multiagent,Krishna2009Auction,Kokott2019Beauty}.

Formally, we model a dynamic auction with $n$ bidders as a partially observable Markov Game $\mathcal{G}=(S,A,R,T,H,o)$, where
\begin{itemize}
        \item $S=\boldsymbol{\Theta} \times \mathcal{H}$ is the state space, where 
        \begin{itemize}
            \item $\boldsymbol{\Theta}=\prod_{i=1}^n \Theta_i$ is the joint type space, and each bidder $i$ has a type in $\Theta_i$.
            \item $\mathcal{H}$ is the history of the auction. After $\tau$ rounds the history contains the past allocations $\{\boldsymbol{x_t}\}_{t\leq \tau}$ and past payments $\{\boldsymbol{p}_t\}_{t\leq \tau}$, as well as other possible contextual information such as which goods are sold currently.
        \end{itemize} 
        \item $\boldsymbol{A}=\prod_{i=1}^n A_i$ is the joint action space of all agents, where $A_i$ denotes bidder $i$'s feasible bids each round.
        \item $T$ is the transition kernel that encodes the mechanism's decisions, i.e. the allocation and payment rule. Note we assume that bidder's types do not change and thus transitions affect only $\mathcal{H}$. More specifically, $$T(s_t,\boldsymbol{a},s_{t+1})=\begin{cases}1 & \text{if } s'=s\cup\{\boldsymbol{x}_t(\boldsymbol{a}), \boldsymbol{p}_t(\boldsymbol{a})\}\\
        0 & \text{else }
        \end{cases}$$
        \item $H$ is the horizon, i.e. the number of rounds played.
        \item $R: S\times A\times S\times \{1,\dots,n\}\times \Theta$ is the reward function that assigns agents their quasi-linear utilities, based on the mechanisms decisions. If the auctioneer uses the assignment function $\boldsymbol{x}$ and payment rule $\boldsymbol{p}$ then $r_{i,t}(s,\boldsymbol{a})=\theta_i x_{i,t}(\boldsymbol{a}) - p_{i,t}(\boldsymbol{a})$.
        \item $o=\prod_{i=1}^n o_i$ are the observations functions of each agent. We assume that $o_i(\boldsymbol{\theta},h)=(\theta_i,h)$. Notice, that we assume perfect recall. What an agent observes once, they will not forget and might use it to gradually learn more about their opponents through explicit or implicit Bayesian updating. The range of $o_i$ is denoted by $\mathcal{O}_i$ and we use ${O}_i$ to denote a given observation of agent $i$.
    \end{itemize}
The auction proceeds as follows: Each round certain goods are sold until the horizon $H$ is reached. Agents make an observation $o_i(s)$ of the current state and bid according to their policies $\pi_i(o_i(s))$. The auctioneer then assigns the goods that were on sale and charges prices according to the assignment and payment rule, which can depend on the history of the game.
Before participating in the game $\mathcal{G}$, each agent samples a type $\theta_i$ from a publicly known distribution $F_i$. These draws together make up $\boldsymbol{\Theta}$. In this work, we assume all agents---except for $i$---play fixed policies, which we denote by $\boldsymbol{\pi}_{-i}$. Note that by fixing $\boldsymbol{\pi}_{-i}$, $i$ thus faces the following MDP
$\mathcal{M}=(O_i,A_i,T_i,H,R_i)$ where
\begin{itemize}
    \item $O_i=(\theta_i,h)$ is the state space of the MDP, which corresponds to the observation space of agent $i$ in the Markov Game.
    \item $A_i$ is the bidding space of agent i.
    \item $T_i$ is the transition kernel $T$ marginalised over the other bidders, i.e. $T_i(O_i,a_i,O'_i)=\mathbb{E}_{\boldsymbol{\theta}_{-i}\sim F_{-i}(O_i)}[\mathbb{E}_{\boldsymbol{a_{-i}}\sim \boldsymbol{\pi}_{-i}}\mathbb{E}_{s'}[T(s,a,s')|o_i(s'_i)=O'_i]]|o_i(s_i)=O_i]$ where $F_i(O_i)$ corresponds to the Bayesian posterior distribution over the opponents type, where $F_i$ is the prior and $O_i$ the observation used to calculate the posterior.
    \item $H$ is the horizon as in the MDP.
    \item $R_i$ is the marginalised reward function of agent $i$, i.e.
    $R_i(O_i,a_i,O'_i)=\mathbb{E}_{\boldsymbol{\theta}_{-i}\sim F_{-i}(O_i)}[\mathbb{E}_{\boldsymbol{a_{-i}}\sim \boldsymbol{\pi}_{-i}}\mathbb{E}_{s'}[R(s,a,s')|o_i(s'_i)=O'_i]]|o_i(s_i)=O_i]$
\end{itemize}

In this setting we study the decision problem of $i$ having to learn a best response to $\boldsymbol{\pi}_{-i}$, i.e. an optimal policy in the induced MDP.

More specifically let 
$$
V_i^{\boldsymbol{\pi}}(O_i)=\mathbb{E}_{s,\boldsymbol{\pi}}[\sum_{t=k}^H r_{i,t}(s_t)|o_i(s_k)=O_i]
$$
be the expected value\footnote{We use the notion of value as used in the RL literature. In the terminology of auction literature we would say expected utility.} in round $k$ for agent $i$ observing $O_i$. Further we say $\pi_i$ is a best response to $\boldsymbol{\pi_{-i}}$, written $\pi_i \in \boldsymbol{BR}(\boldsymbol{\pi}_{-i},O_i)$ if
$$
V_i^{\pi_i,\boldsymbol{\pi_{-i}}}(O_i) \geq V_i^{\pi'_i,\boldsymbol{\pi_{-i}}}(O_i) \quad \forall \pi'_i \in \Pi_i
$$
where $\Pi_i$ be the policy space of $i$. 
Our aim is to find an optimal policy, such that 
$$
\pi_i^*(O_i) \in \boldsymbol{BR}(\boldsymbol{\pi}_{-i},O_i) \quad \forall O_i
$$

\section{Challenges in using RL for auctions}
Learning in auctions is hard for two reasons: on the one hand, agents have incomplete information and thus need to remember and reason over all of their past observations. On the other hand, auctions are highly stochastic environments where the outcome of a given policy can be very different depending on the types drawn by all agents.

To model dynamic auctions as MDPs, we thus need to include bidders' types, as well as the history of the auction, in the state space. To account for the randomness and the incomplete information, the MDP's state space becomes very large even for simple auction settings. To better understand and solve this problem we distinguish between observable and unobservable states.

\begin{itemize}
    \item \textbf{Observable states} At the beginning of the auction, agents only observe their own type, which stays the same throughout the game. Thus, in order for $i$ to learn an optimal policy for every state, $\mathcal{G}$ needs to be initialised many times with different types. In essence, $i$ is learning different policies for slightly different "games", parameterised by its own type. Notice however, that the mechanics of the game, i.e. the action space and transition function do not change. $\theta_i$ only has an influence on $i$'s rewards $r_i$. This paradigm is well-known in reinforcement learning and robotics under the name multi-task RL \citep{Eysenbach2020Rewriting}. In such a setting, it can be beneficial for the agent learning a task $\theta_i$ to use experience gathered as part of other tasks as well. Indeed if the transitions are known, the agent can easily \textit{relabel} its experience (i.e. state, action, state transitions) for another task by recomputing the reward function $r_i(\cdot|\theta_i)$. Intuitively this can be helpful as a certain action might be unsuitable for the task as part of which it was explored, but optimal for another one. We therefore propose to use experience relabelling to learn more efficiently.
    
    \item \textbf{Unobservable states} Keeping both $\theta_i$, as well as a deterministic policy $\boldsymbol{\pi}$ fixed, the possible transitions in $\mathcal{G}$ can still have support on an uncountable set of states with the possible rewards $r_i$ having a large variance. This is because of the randomness in the opponents types $\theta_{-i}$. From the point of agent $i$ we have to assume this is part of the stochasticity of the environment. 
    
\end{itemize}

\section{Methods}
In this work we use RL to learn an optimal policy for one agent participating in an auction against fixed opponents. We have several desiderata for our learning algorithm. These include:
\begin{itemize}
    \item \textbf{Model-free learning} For transparency reasons, agents generally understand the auction mechanism that they are participating in. However, we don't want to assume we have a complete model of the environment (writing down all states, transitions, calculating correct Bayesian updates etc.) -- only the ability to query a simulator of the auction. Thus we consider model-free RL methods which can learn in such settings given only sampled environment trajectories and a reward signal. In principle, model-based RL techniques involving planning over a learned model of the environment could also be used, but for the time being we stick to model-free methods.
    \item \textbf{Experience Relabeling}
    As discussed above, we propose to use experience relabelling to learn the different tasks induced by the different types of the agent more efficiently. Using the transitions sampled from policies for other agent types, we therefore can ensure sufficient exploration of the policy space for a given agent type.
    \item \textbf{Off-policy learning} If agents relabel their experiences, this changes the distribution of the training data. It is therefore best used in combination with off-policy RL methods, that do not necessitate the training data to be generated by the current policy. 
    \item \textbf{Function approximation} To cope with the large state space we want to approximate the policy and value functions using neural networks.

\end{itemize}
The above considerations lead us to employ Soft Actor-Critic (SAC), a variant of the actor-critic algorithm with entropy regularisation \citep{Haarnoja2018Soft}.

In maximum entropy RL, the objective is to maximise the expected sum of rewards---as standard for any RL problem---plus  an entropy regularisation term to ensure policies remain non-deterministic.
\begin{equation}
\label{eq:obj}
J(\pi)=\sum_{t=0}^H \mathbb{E}_{\left(\mathbf{s}_t, \mathbf{a}_t\right) \sim {\pi}}\left[r\left(\mathbf{s}_t, \mathbf{a}_t\right)+\alpha \mathcal{H}\left(\pi\left(\cdot \mid \mathbf{s}_t\right)\right)\right]
\end{equation}
Recall that the Q-function of a policy can be written via the Bellman equation as
$$
q_{\pi}(s,a)=r(s,a) + \mathbb{E}_{s'|(s,a),a'\sim \pi(|s')}[q_{\pi}(s',a')]
$$
A parameterised approximation of this function for non-deterministic policies can be learned by minimising the following loss function with respect to $\boldsymbol{\theta}_Q$ using stochastic gradient descent.
$$
\frac{1}{2} \mathbb{E}_{a^{\prime} \sim \pi}\left[r(s,a)+\gamma Q\left(s^{\prime}, \boldsymbol{a}^{\prime} ; \boldsymbol{\theta}_{\mathrm{Q}}^{\text {old }}\right)-Q\left(s, a ; \boldsymbol{\theta}_{\mathrm{Q}}\right)\right]^2
$$
Based on the learned Q-function \footnote{Or more precisely the soft Q-function, which is the Q-function with an added entropy regularisation term.}, SAC improves the policy by minimising the KL-divergence to the soft-max of the learned Q-values
$$
\pi_{\text {new }}=\arg \min _{\pi^{\prime} \in \Pi} \mathrm{D}_{\mathrm{KL}}\left(\pi^{\prime}\left(\cdot \mid \mathbf{s}_t\right) \| \frac{\exp \left(Q^{\pi_{\mathrm{old}}}\left(\mathbf{s}_t, \cdot\right)\right)}{Z^{\pi_{\mathrm{old}}}\left(\mathbf{s}_t\right)}\right)
$$

Note that if the optimal Q-function is known, minimising the KL-divergence is equivalent to maximising the objective in Equation \ref{eq:obj} \citep{Haarnoja2018Soft}. The pseudocode of SAC can be found in Algorithm \ref{alg:SAC}. In our proposed learning approach (see Algorithm \ref{alg:pseudolearner}), we combine SAC with experience relabelling---a technique from multi-task RL \citep{Eysenbach2020Rewriting}.

\begin{algorithm}
\KwIn{$B,\phi,\psi^{(1)},\psi^{(2)}$} 
 For each transition in $B$ compute target\\
 $y= r + \gamma \min(Q_{\psi^{(1)}}(s_{t+1},\pi_{\phi}(O_{t+1})),Q_{\psi^{(2)}}(O_{t+1},\pi_{\phi}(O_{t+1})))$ \tcc*{Using double Q learning}
 \tcc{Update critic}
 \tcc{Gradient steps are actually made using ADAM, but simplified here}
 $\psi^{(1)}\gets \psi^{(1)}-\eta_1 \nabla_{\psi^{(1)}} \frac{1}{B}\sum_{(s,a,r,s',y)\in B}(Q_{\psi^{(1)}}(O_t,a_t)-y)^2$\\
 $\psi^{(2)}\gets \psi^{(2)}-\eta_1 \nabla_{\psi^{(2)}} \frac{1}{B}\sum_{(s,a,r,s',y)\in B}(Q_{\psi^{(2)}}(O_t,a_t)-y)^2$\\
 \tcc{Update Actor}
 \tcc{Gradient computed with reparametrization trick}
 $\phi \gets \phi - \eta_2 \nabla_{\phi} \mathbb{E}_{O_t\sim \mathcal{D},a_t\sim \pi_{\phi}}[\log(\pi_{\phi}(a_t|O_t))-\min(Q_{\psi^{(1)}}(O_{t},a_t),Q_{\psi^{(2)}}(O_{t},a_t))] $\\
 \KwRet{$\phi,\psi^{(1)},\psi^{(2)}$}
 \caption{Pseudocode of SAC}\label{alg:SAC}
\end{algorithm}

\SetKwComment{Comment}{/* }{ */}
\begin{algorithm}
 initialize $\phi,\psi^{(1)},\psi^{(2)},j \gets 0$ \;
 \While{$j \leq$ numEpochs}{
  \For{$k=1$ \KwTo numExperience}
{  
$\theta_i \sim F_i$ \tcc*{Sample type}
$O_0 =({\theta_i},\emptyset)$\tcc*{Initialise observation with empty history}
    \For{$t=0$ \KwTo $H$}{
    $\mathbf{a}_t \sim \pi_\phi\left(\mathbf{a}_t \mid O_t\right)$ \\
    $O_{t+1} \sim p\left(O_{t+1} \mid O_t, \mathbf{a}_t\right)$\\
    $\mathcal{D} \leftarrow \mathcal{D} \cup\left\{\left(O_t, \mathbf{a}_t, r\left(O_t, \mathbf{a}_t\right), O_{t+1}\right)\right\}$\\
    }
}
Sample minibatch $B$ from $\mathcal{D}$ using experience relabelling \\
$\phi,\psi^{(1)},\psi^{(2)} \gets \texttt{SAC}(B,\phi,\psi^{(1)},\psi^{(2)})$  \\
$j \gets j+1$
}
\KwRet{$\pi_{\phi}$}\;

\caption{Pseudocode of our Learning Algorithm}\label{alg:pseudolearner}
\end{algorithm}

\section{Experiments}

\subsection{Environments}
We begin by describing the analytical equilibria of the auction environments studied.
\subsubsection{Sequential Sales environment}
\label{sec:brseqsal}
\citet{Krishna2009Auction} discusses sequential auctions where a single unit of a good is sold in each round. Bidders have unit demand, i.e. if a bidder wins a good, they leave the auction, and a fixed welfare for gaining the good.

In this environment we distinguish between different settings. On the one hand we study two pricing rules: either first price, i.e. highest bidder pays his bid, or second price, i.e. highest bidder pays second highest bid. Moreover, we distinguish by the behaviour of the fixed opponents, either they bid truthfully or they bid according to the known equilibrium policies.

\paragraph{Equilibrium strategies}
The equilibrium for a first-price sequential sales auction with $N$ bidders and $K$ items with types uniform on $[0,1]$ is $\pi_i^*(\theta_i) = \frac{N - K}{N - k + 1} \theta_i$ in the $k$th round. For a second price auction the equilibrium policy is $\pi_i^*(\theta_i) = \frac{N - K}{N - k} \theta_i$ \citep{Krishna2009Auction}.

\paragraph{Best-responses to truthful bidding}

If opponents bid truthfully, the optimal strategy for bidder $i$ involves bidding 0, i.e. losing, until the last round. At this point, the observed sale prices provide information on the types of the bidders who have not yet exited the auction. For a second-price auction, the best policy is then to bid $\min(\theta_i,p_{min})$, where $p_{min}$ is lowest price observed so far; for first-price auctions, the optimal policy is $\min(\theta_i/2,p_{min})$.

\subsubsection{Split-award environment}
\cite{Kokott2019Beauty} present a combinatorial split award auction, used in procurement. It therefore is a \textit{reverse} auction, which means bidders are not buying something from the auctioneer but selling, and the auctioneer wants to buy something at a low price. There are two units, which the auctioneer want to procure. He is indifferent whether they come from two sellers or only one. The auction has a maximum of two phases. In the first round, bidders submit a bid (price offer) for both the \textit{sole} (both units) and the \textit{split} (one unit) award. If the unit price of the lowest split bid is higher than the unit price of the lowest sole bid, the auction is over and the bidder with the lowest sole bid receives that amount of money for both units.

Otherwise, the bidders with the lowest split offer wins and receives his bid. Then the auction moves to a second round where bidders can continue bidding prices at which they are willing to sell the second unit.

Before the auction the bidders draw their type $\theta \in [\underline{\Theta},\overline{\Theta}]$ according to a distribution $F$, which determines their cost for the sole award. Moreover, there is a publicly known efficiency/scale parameter $C$ that determines whether there are economies of scale or diseconomies of scale. Specifically, the cost of bidder $i$ for the first split is $C\theta_i$, whereas the cost of the second split is $(1-C)\theta_i$. In this work we have so far focused on the setting of \textit{Dual-Source efficiency} (DSE), which means that the diseconomies of scale are so strong that procuring the goods from two different bidders is always more efficient than buying it from one. This is the case when $C\leq \frac{\underline{\Theta}}{2\overline{\Theta}}$ 

Again we distinguish between bidding against equilibrium strategies and bidding against truthful bidders.

\paragraph{Equilibrium strategies}
\cite{Kokott2019Beauty} present the equilibrium in this environment for $n>3$ bidders as
$$
\begin{aligned}
p_e^{\sigma 1}\left(\theta_i, h^0\right) & =\frac{\int_{\theta_i}^{\bar{\Theta}} p_e^{\sigma 2 l}(t)(n-1)(1-F(t))^{n-2} f(t) d t}{\left(1-F\left(\theta_i\right)\right)^{n-1}} \\
p_e^{\sigma 2 w}\left(\theta_w, h^1\right) & =\theta_w(1-C) \\
p_e^{\sigma 2 l}\left(\theta_l, h^1\right) & =\theta_l C+C \frac{\int_{\theta_l}^{\bar{\Theta}}(1-F(t))^{n-2} d t}{\left(1-F\left(\theta_l\right)\right)^{n-2}}
\end{aligned}
$$
where $p_e^{\sigma 1}$ is the split bid in the first round and $p_e^{\sigma 2 w}$ is the split bid of the first round winner $w$, whereas $p_e^{\sigma 2 l}$ is the split bid of the looser $l$. Note there is no unique optimal sole bid. Instead any bid high enough to avoid winning the sole award is sufficient.
\paragraph{Best-responses to truthful bidding}
In this environment, we consider the setting when there are two agents in total and the second one is bidding truthfully. As the learning agent is playing against a truthful opponent and the setting is DSE, he can never win the sole award without incurring a loss. Hence the best strategy is trying to win a split-award. Similarly to the sequential sales setting, the best strategy is to lose the first round as this gives the agent information about the exact type $\theta_o$ of its opponent. Moreover, the opponent is less competitive in the second round due to the diseconomies of scale.\footnote{The first round it bids $C\theta_o$, which is always lower than the second round bid $(1-C)\theta_o$ if the opponent won.} In the second round the agent should then bid exactly $(1-C)\theta_o$ to achieve maximum utility.\footnote{This is optimal because \cite{Kokott2019Beauty} assume that the tiebreak decides in favour of splitting the goods between two bidders.}
\subsection{Results}
Now let us turn to our experimental results. In our experiments unless otherwise mentioned, the Q-function and the policy are approximated using neural networks with 2 hidden layers and 256 units per hidden layer using ReLU as activation function.
We use a modified version of the SAC implementation from~\cite{geng2022jaxcql}.
The policy network learns a mean and variance, which we map to the bids $a_t$ as follows
    \begin{align*}
        m,\sigma & = \texttt{NN}_{\pi}(O_t)\\
        x & \sim \mathcal{N}(m,\sigma) \\
        a_t &= \frac{\theta_{max} - \theta_{min}}{2}  \tanh(x) + \frac{\theta_{max} + \theta_{min}}{2}
    \end{align*}

To judge the learned policies, we measure their $\ell^2$ distance to the optimal policies. Moreover, we measure the difference between average reward achieved vs. maximum possible and  all via Monte Carlo estimates over 4000 sampled type profiles.

\subsubsection{Split-award, truthful bids}
We consider the setting with
\begin{itemize}
    \item $n=2$ bidders
    \item $[\underline{\Theta},\overline{\Theta}]=[1,2]$
    \item $C=0.2$ (This ensures Dual Source Efficiency, i.e. the split-award is always the efficient outcome.)
\end{itemize}
In this setting, the optimal policy described earlier corresponds to bidding above 0.4 and 0.8 for the split respectively sole award in the first round to ensure loosing an then bidding $0.8\theta_o$ in the second round. 

We train for 3000 epochs on 500 parallel environments with learning rates of $10^{-3}$ and $3 \times 10^{-3}$ for the policy and critic networks respectively, with a target entropy of $-20$.

As shown in Table \ref{tab:results_split} the learned policy is very close to optimal. The agent loses in the first round and just slightly underbids the opponent in the second round. Moreover, the learned value function of the critic is close in $\ell_2$ distance to the true value function computed analytically for the optimal policy.

\subsubsection{Split-award, equilibrium bids}
We study the following variant of this auction.
\begin{itemize}
    \item $n=3$ bidders
    \item $[\underline{\Theta},\overline{\Theta}]=[1,2]$
    \item $C=0.2$
\end{itemize}
As the learning agent is playing against equilibrium bids, it is expected that it learns the equilibrium policies or at least policies with the same expected value.\footnote{Consider the example of an agent playing Rock Paper Scissors against a fixed opponent playing all three actions with equal probability. As the opponent is fixed, any strategy has the same expected value of 0 and would thus be a best response.} In the setting studied here, the equilibrium policy presented earlier translates to an optimal first round split bid of
$$
\frac{0.2}{3}(\theta_i+4)
$$
and an optimal second round split bid of
$$
0.2\left(\theta_i+\frac{2-\theta_i}{2}\right)
$$
For a derivation of these results, we refer the reader to Appendix \ref{sec:kokott}.

We use the same hyperparameters as for the truthful split-award auction. In our experiments, as shown in Table \ref{tab:results_split} the agent learns a policy with almost the same expected utility as in equilibrium, as well as a low $\ell_2$ distance to equilibrium policies.

\begin{table}[]
\centering

\caption{Summary of runs in the split-award environment with distances to optimal policies and values.}
\begin{tabular}{lll}
\toprule
\textbf{Bidders}                          & \multicolumn{1}{c}{2} & \multicolumn{1}{c}{3} \\
\textbf{Environment}                      & split                 & split                 \\
\textbf{Opponent Strategy}                & truthful              & equilibrium           \\
\midrule
{Optimal Reward}                  & 0.9                   & 0.05                  \\
{Achieved Reward}                 & 0.8813                & 0.04981               \\
{Utility Difference} & 0.0187 &0.00019 \\
{$\ell^2$ First Round Split}             & 0                     & 0.0126                \\
{$\ell^2$ First Round Sole}              & 0                     & 0                     \\
{$\ell^2$ Second Round}                  & 0.018                 & 0.0026                \\
{$\ell^2$ First Round Value Function}    & 0.013                 & 0.0009                \\
\bottomrule
\end{tabular}
\label{tab:results_split}
\end{table}

\subsection{Sequential Sales}

\subsubsection{First price, 1 good}
We consider the following setting 
\begin{itemize}
    \item 1 item sold in 1 round
    \item 2 bidders
    \item Highest bidder wins and pays its bid
    \item Second bidder bids the equilibrium bid of $\frac{1}{2}\theta_2$
\end{itemize}

We train with the same parameters as for the split-award setting, but with a target entropy of $-10$. 

In this setting it is well-known that the equilibrium is bidding half the true type \citep{Krishna2009Auction} and playing against an equilibrium bidder, we thus expect the learning agent to converge to this strategy. Indeed as shown in Table \ref{tab:results_seq} this is exactly the case.

\subsubsection{Second price, 1 good}
We consider the following setting:
\begin{itemize}
    \item 1 item sold in 1 round
    \item 2 bidders
    \item Highest bidder wins and pays bid of second highest bidder.
    \item Second bidder bids the equilibrium bid of $\theta_2$
\end{itemize}

We train with the same parameters as for the split-award setting, but with a target entropy of $-10$. 

This is the famous second-price auction, where it is well-known that bidding your true type is optimal, and playing against an equilibrium bidder, we thus expect the learning agent to converge to this strategy. 
Indeed as shown in Table \ref{tab:results_seq} this is the case.
\begin{figure}
    \centering
    \includegraphics[width=0.4\textwidth]{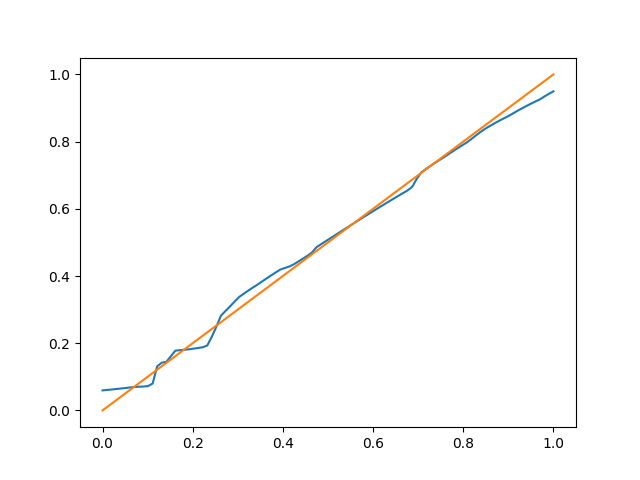}
    \includegraphics[width=0.4\textwidth]{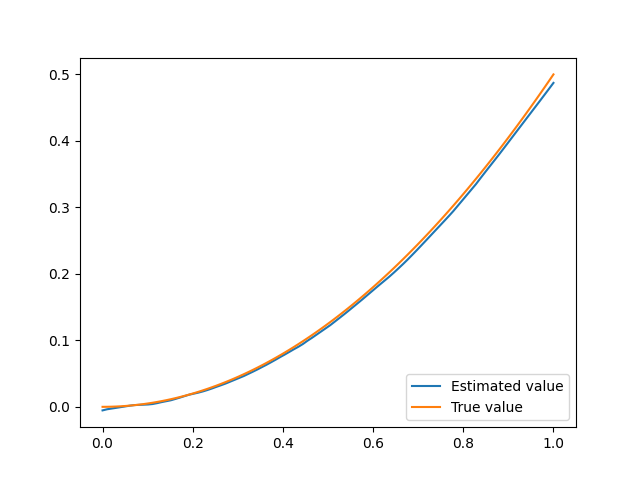}
    \caption{Bids (left) and expected utilities (right) in one round second price auction. Blue is the learned policy/learned value under learned policy, and orange the true best response/true value. Small deviations from the optimal policy result result in very small changes to the reward. Similar ``noise'' is observed in other methods for learning bids, e.g. \cite{Fichtl2022Computing}.}
    \label{fig:seqbid}
\end{figure}

\begin{figure}
    \centering
    \includegraphics[width=0.7\textwidth]{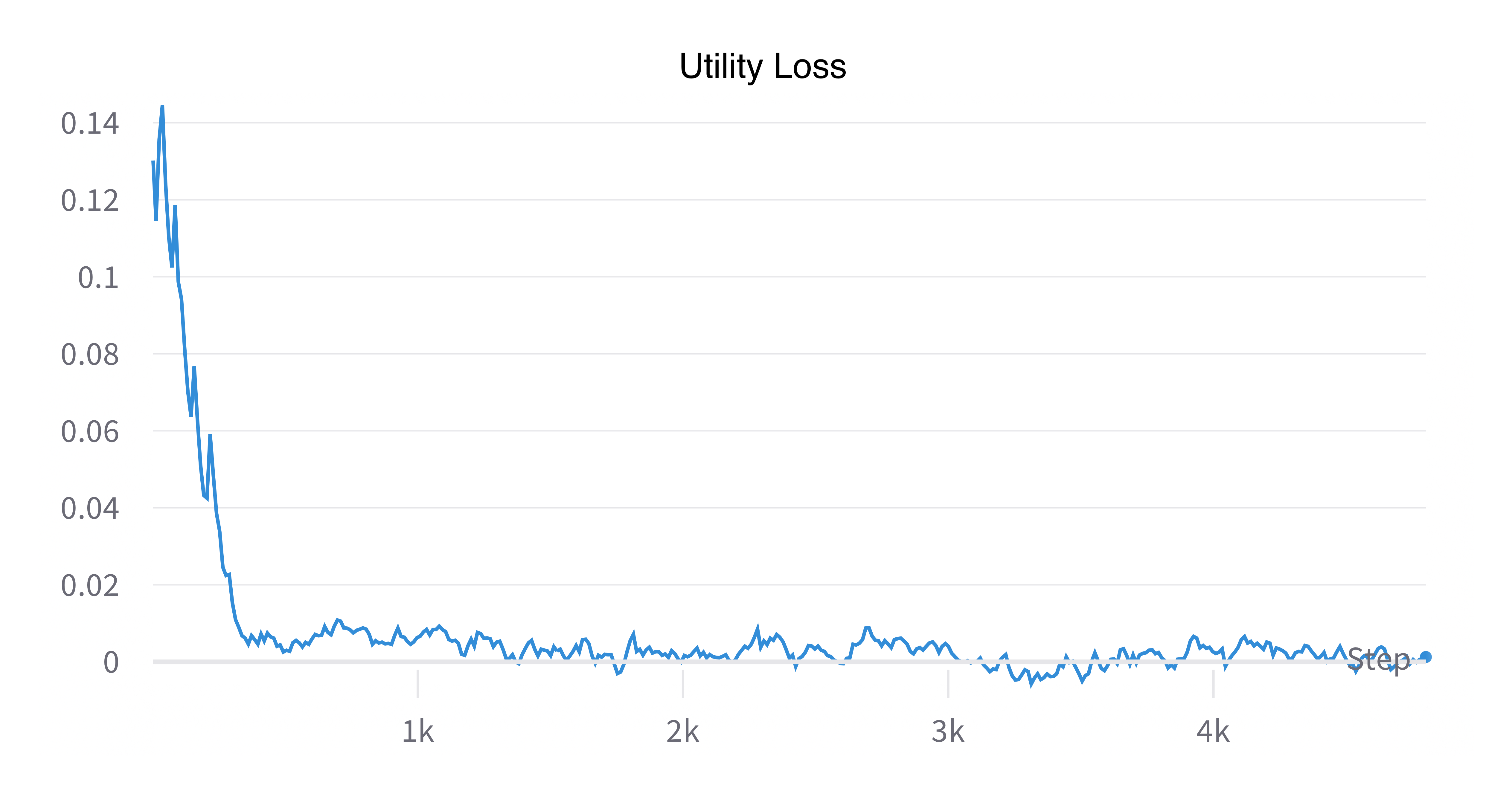}

    \caption{Estimated utility loss achieved by learner in one round second price auction compared to optimal response. Note that the estimated loss is sometimes negative because of the inaccuracy from using Monte Carlo estimation.}
    \label{fig:seqret}
\end{figure}

\subsubsection{First price, 2 goods}
We consider the following setting 
\begin{itemize}
    \item 2 items sold in 2 rounds, 1 item in each round.
    \item 3 bidders.
    \item Highest bidder wins, pays his bid and leaves the auction.
\end{itemize}

\begin{figure}
    \centering
    \includegraphics[width=0.7\textwidth]{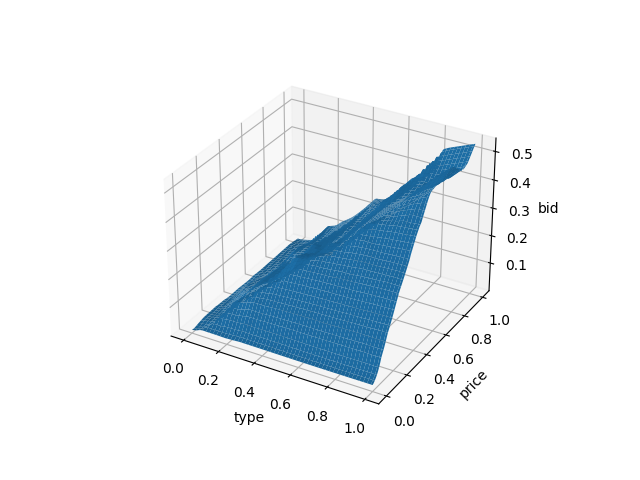}
    \caption{Bidding strategy, given own type and observed price, for the second round of a two-good sequential sales auction under the first price rule. The learned function is very close to the correct best response of bidding $\min(\theta_i/2,p_{min})$.}
    \label{fig:3x2fp_bidding_2ndround}
\end{figure}

\paragraph{Truthful opponents}
We first consider truthful opponents. We use 500 parallel environment copies, training for 3000 epochs with 200 training update steps per epoch. We have a policy learning rate of $10^{-4}$ and a q-function learning rate of $3 \times 10^{-4}$, with a target entropy of $-10$. In this setting, as discussed in Section \ref{sec:brseqsal} the best response strategy is to lose until the last round, and then bid $\min(\theta_i/2,p_{min})$. As shown in Table \ref{tab:results_seq} our algorithm learns a policy that is similar to the optimal one described here, but further away than the policies learned in other settings. However, the utility difference is close to 0, indicating good learning performance. The learned best response in the second round is shown in Figure \ref{fig:3x2fp_bidding_2ndround}

\paragraph{Equilibrium opponents}

\begin{figure}
    \centering
    \includegraphics[width=0.7\textwidth]{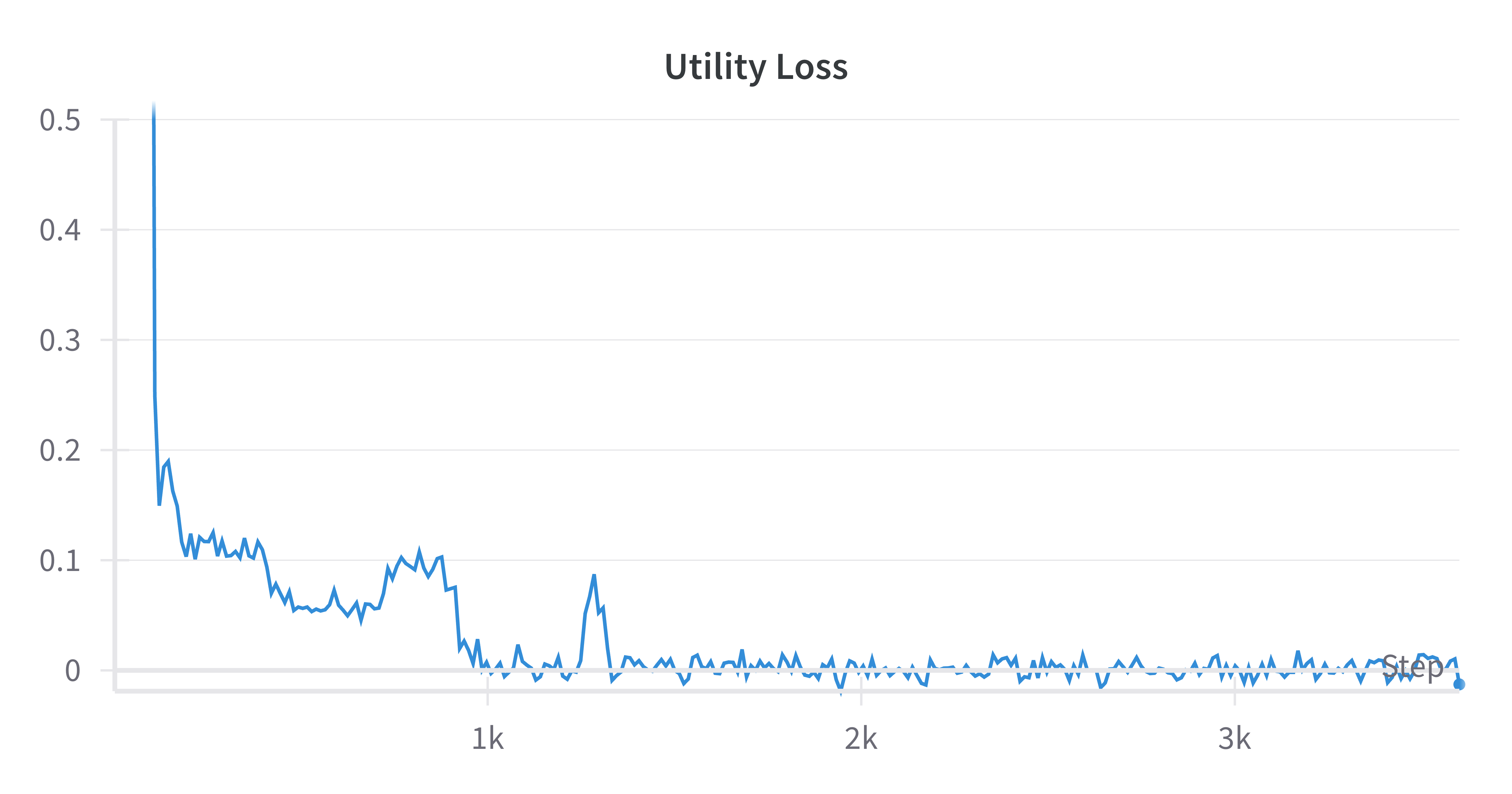}

    \caption{Estimated utility loss achieved by learner in a two-round auction against two other equilibrium bidders, compared to the equilibrium utility. Note that the estimated loss is sometimes negative because of the inaccuracy from using Monte Carlo estimation.}
    \label{fig:seqret2}
\end{figure}

We can also consider the symmetric equilibrium strategy for this setting. In this case, we use a policy network with a single hidden layer of size 64, a critic network with 2 hidden layers of size 64, learning rates of $3 \times 10^{-3}$, and an entropy target of -5. For this setting, we remove the activation function on the final layer, instead simply penalizing negative bids. Again we train for 3000 epochs with 200 RL update steps per epoch. Again, as shown in Table \ref{tab:results_seq} the $\ell_2$ errors of the strategies become relatively small, and the utility of the learned policy is quite close to the equilibrium reward of 0.25 (slightly above due to inaccuracies from MC estimation), as also shown in figure \ref{fig:seqret2}.

\begin{table}
    \caption{Results from experiments with sequential sales environment, reported based on the Monte Carlo estimate calculated in the last round.}
    \centering
\begin{tabular}{lrrrr}
\toprule
\textbf{Bidders}           & 2            & 2           & 3    & 3       \\
\textbf{Rounds}            & 1            & 1           & 2    & 2       \\
\textbf{Payment rule}      & Second price & First price & First price & First price \\
\textbf{Opponents}         & Equilibrium  & Equilibrium & Truthful   & Equilibrium \\
\midrule
{Optimal Strategy Reward}   & 1/6 (exact)         & 1/6 (exact)        & 0.1458  &  .25   \\
{Achieved Reward} & 0.1654       & 0.1657      & 0.1442  &  .262   \\
{Utility Difference}            & 0.0013       & 0.0010      & 0.0016 &  .002    \\
{$\ell^2$ First Round}      & 0.0191       & 0.0087      & 0.0528  &   .03  \\
{$\ell^2$ Second Round}         & -            & -           & 0.0311  &  .01  \\
\bottomrule
\end{tabular}
    \label{tab:results_seq}
\end{table}

\section{Conclusion}
Many dynamic auction formats used today are too complex for either the designers or the bidders to understand their equilibria and bid accordingly. Thus in this work we propose to model the problem of an individual agent learning to bid against a fixed population of bidders as an MDP, and learn to bid using deep RL. By using soft actor-critic, a method which involves a replay buffer, we are able to treat the bidding game as a multi-task problem parameterized by bidder types, allowing us to relabel previously-existing transitions with new types. 

Using our techniques, we are able to reproduce known equilibria in multiple dynamic settings \citep{Kokott2019Beauty,Krishna2009Auction}, as well as find profitable deviations when other bidders play exploitable strategies.
Directions for further work include further scaling of our method, experiments with richer relabeling strategies (e.g.~\cite{Eysenbach2020Rewriting}), extensions to more complex bidder utility models, and extensions to richer auction domains such as dynamic combinatorial auctions with value and demand queries.

\bibliography{references}
\bibliographystyle{tmlr}

\appendix

\section{Re-derived split-award equilibrium}
\label{sec:kokott}
First let us translate the results from the paper to our setting.
We consider the following:
\begin{itemize}
    \item $\underline{\Theta} =1$
    \item $\overline{\Theta}=2$
    \item $n = 3$
    \item $F(t) = t-1$ (uniform on [1,2])
\end{itemize}
In the case of dual source efficiency (DSE) \footnote{This is defined as a setting of diseconomies of scale where it is always efficient to award two splits. For 3 or more bidders this should be the case when $C \leq \frac{\underline{\Theta}}{2 \overline{\Theta}}$, which in our setting means $C\leq 0.25$. As we choose $C = 0.2$ this should be fine.}
$$
\begin{aligned}
p_e^{\sigma 1}\left(\Theta_i, h^0\right) & =\frac{\int_{\Theta_i}^{\bar{\Theta}} p_e^{\sigma 2 l}(t)(n-1)(1-F(t))^{n-2} f(t) d t}{\left(1-F\left(\Theta_i\right)\right)^{n-1}} \\
p_e^{\sigma 2 w}\left(\Theta_w, h^1\right) & =\Theta_w(1-C) \\
p_e^{\sigma 2 l}\left(\Theta_l, h^1\right) & =\Theta_l C+C \frac{\int_{\Theta_l}^{\bar{\Theta}}(1-F(t))^{n-2} d t}{\left(1-F\left(\Theta_l\right)\right)^{n-2}}
\end{aligned}
$$
Moreover, the sole award bid should be high enough so that the auctioneer always awards a split-award.
The second round bidding strategy in this case is:
$$
\Theta_w(1-C)=\Theta_w 0.8
$$
The second round losing strategy is given by:
$$
\begin{aligned}
    p_e^{\sigma 2 l}\left(\Theta_l, h^1\right) & =\Theta_l C+C \frac{\int_{\Theta_l}^{\bar{\Theta}}(1-F(t))^{n-2} d t}{\left(1-F\left(\Theta_l\right)\right)^{n-2}}\\
    &= \Theta_l C+C \frac{\int_{\Theta_l}^{\bar{\Theta}}(2-t)d t}{(2-\Theta_l )}\\
    &= \Theta_l C+C \frac{2t - t^2/2 \big |_{\Theta_l}^{2}}{(2-\Theta_l )}\\
    &= C(\Theta_l + \frac{2-2\Theta_l + \frac{1}{2}\Theta^2_l}{(2-\Theta_l )})\\
    &= C(\Theta_l + \frac{(2-\Theta_l)^2}{2(2-\Theta_l )})\\
    &= C(\Theta_l + \frac{(2-\Theta_l)}{2})\\
\end{aligned}
$$
Finally let us move to the first round split strategy, where we plugin the second round split strategy

$$
\begin{aligned}
p_e^{\sigma 1}\left(\Theta_i, h^0\right) & =\frac{\int_{\Theta_i}^{\bar{\Theta}} p_e^{\sigma 2 l}(t)(n-1)(1-F(t))^{n-2} f(t) d t}{\left(1-F\left(\Theta_i\right)\right)^{n-1}} \\ 
&= 2\frac{\int_{\Theta_i}^{\bar{\Theta}} C(t + \frac{(2-t)}{2})(2-t)d t}{(2-\Theta_i)^2} \\
&= 2\frac{\int_{\Theta_i}^{2} C(t + \frac{(2-t)}{2})(2-t)d t}{(2-\Theta_i)^2}
\end{aligned}
$$

This integral evaluates to:
$$
\begin{aligned}
p_e^{\sigma 1}\left(\Theta_i, h^0\right) & = 2\frac{\int_{\Theta_i}^{2} C*(t + \frac{(2-t)}{2})(2-t)d t}{(2-\Theta_i)^2}\\
&= C\frac{1/3(2-\Theta_i)^2(\Theta_i+4)}{(2-\Theta_i)^2}\\
&= 1/3(\Theta_i+4)C
\end{aligned}
$$

Last but not least, the expected utility of the optimal policy in the first round, having observed $\theta_i$ is given by
$$(0.2/3(x+4)-0.2x)(2-x)^2+ (x0.2+0.2(2-x)/2-0.2x)x-1)(2-x)2.$$

\end{document}